\documentstyle[11pt]{article}
\input epsf

\makeatletter

\@addtoreset{equation}{section}
\makeatother

\begin{document}
\baselineskip 20pt
\rightline{CU-TP-739}
\rightline{hep-th/9602167}
\vskip 2cm
\centerline{\Large\bf The Moduli Space of Many BPS Monopoles}
\centerline{\Large\bf for Arbitrary Gauge Groups}
\vskip 1cm
\centerline{\large\it
Kimyeong Lee,\footnote{electronic mail: klee@phys.columbia.edu}
Erick J. Weinberg,\footnote{electronic mail: ejw@phys.columbia.edu} and 
Piljin Yi\footnote{electronic mail: piljin@phys.columbia.edu}}
\vskip 1mm
\centerline{Physics Department, Columbia University, New York, NY 10027}
\vskip 2cm
\centerline{\bf ABSTRACT}
\vskip 1cm
\begin{quote}

We study the moduli space for an arbitrary number of BPS monopoles 
in a gauge theory with an arbitrary gauge group that is maximally 
broken to $U(1)^k$. From the low energy dynamics of well-separated 
dyons we infer the asymptotic form of the metric for the moduli space.  
For a pair of distinct fundamental monopoles, the space thus obtained 
is $R^3 \times(R^1\times {\cal M}_0)/Z$ where ${\cal M}_0$ is the 
Euclidean Taub-NUT manifold.  Following the methods of Atiyah and 
Hitchin, we demonstrate that this is actually the exact moduli space 
for this case.  For any number of such objects, we show that the 
asymptotic form remains nonsingular for all values of the intermonopole 
distances and that it has the symmetries and other characteristics 
required of the exact metric. We therefore conjecture that the 
asymptotic form is exact for these cases also.
\end{quote}

\newpage
\setcounter{footnote}{0}
\section{Introduction}

Among the many remarkable features of the Bogomol'nyi-Prasad-Sommerfield
(BPS) limit \cite{bps} is the existence of families of degenerate static
multimonopole solutions.  For any given topological charge, the
space of such solutions, with gauge equivalent configurations identified,
forms a finite dimensional moduli space.   A metric for this space is
defined in a natural way by the kinetic energy terms of the Yang-Mills-Higgs 
Lagrangian.  As was first pointed out by Manton \cite{geodesic}, a
knowledge of this metric is sufficient for determining the low energy
dynamics of a set of monopoles and dyons.  More recently [3-5], such knowledge
has been used in theories with extended supersymmetry to show the
existence of some of the dyonic states required by the 
electromagnetic duality conjecture of Montonen and Olive \cite{dual}. 

For the case of an $SU(2)$ gauge symmetry spontaneously broken to
$U(1)$, the moduli space ${\cal M}$ of solutions carrying $n$ units
of magnetic charge has $4n$ dimensions.  This naturally suggests that
these solutions should be interpreted as configurations of $n$
monopoles, each of which is specified by three position coordinates
and a $U(1)$ phase angle.  (Recall that time variation of this phase
angle gives an electrically charged dyon.)  The metric for the
two-monopole moduli space was determined by Atiyah and Hitchin
\cite{atiyah}.  For three or more monopoles, the moduli space metric
is still unknown.  An asymptotic form, valid in the regions of ${\cal M}$
corresponding to widely separated monopoles, has been found by Gibbons
and Manton \cite{gary}, but this develops singularities if any of the
intermonopole distances becomes too small, and hence cannot be exact.

In this paper, we consider an arbitrary gauge group $G$ with rank $k
\ge 2$, with the adjoint Higgs field assumed to be such as to give the
maximal symmetry breaking, to the Cartan subgroup $U(1)^k$.  In this
case, there are $k$ independent topological charges $n_a$ and,
correspondingly, $k$ ``fundamental monopoles'', each of which carries
a single unit of one of these charges \cite{erick}.  The moduli spaces
corresponding to any combination of $n$ fundamental monopoles are
$4n$-dimensional, just as in $SU(2)$.  We determine the moduli space
metric for all two-monopole solutions; this generalizes the $SU(3)$
result outlined by us in \cite{klee} and also found by Gauntlett and
Lowe \cite{gaunt} and by Connell \cite{connell}.  In addition, we find
the asymptotic 
form of the metric for any number of monopoles.  We then argue that in
a number of cases with three or more monopoles this asymptotic form is
in fact likely to be the exact metric over the entire moduli space.

It might seem odd to try to obtain the moduli space metric for three
or more monopoles with a larger gauge group when this cannot even be
done for $SU(2)$.  The reason that it might actually be easier to work
with a larger group is that in a number of such cases the moduli space
possesses greater symmetry.  Specifically, consider a multimonopole
solution with $n \le k$ fundamental monopoles, each corresponding to a
different topological charge.  The corresponding $U(1)$ factors of the
unbroken gauge group act nontrivially on this solution, and this
action generates an isometry on the moduli space.  Furthermore, for
any such collection of topological charges there is a solution that,
although it is composite, is spherically symmetric.  On the moduli
space, such a solution corresponds to a fixed point under the
isometries that correspond to overall rotation of the monopole
configuration.  In $SU(2)$ there are no spherically symmetric
solutions with multiple magnetic charge \cite{guth}, and hence no such
fixed points.

We recall some properties of BPS monopoles in an $SU(2)$ theory in
Sec.~2, and then discuss the solution for higher groups in Sec.~3.  We
then begin our analysis of the moduli spaces in Sec.~4.  Here,
following the approach that Gibbons and Manton used for $SU(2)$, we
use our knowledge of the interactions between widely separated dyons
to infer the asymptotic form of the moduli space metric.  In Sec.~5,
we specialize to the case of two monopoles.  After separating out the
center-of-mass motion, we are left with a relative moduli space ${\cal
M}_0$ that is simply a Taub-NUT manifold.  Although our methods thus
far only establish that this is the correct manifold asymptotically,
we note that it remains smooth as the monopole separation is taken to
zero and, further, that it possesses the appropriate symmetries.  This
suggests that this asymptotic form might in fact be correct.  In
Sec.~6 we use the methods of Atiyah and Hitchin to verify this.  The
argument here rests on the fact that the moduli space must be
hyperk\"ahler.  The Taub-NUT space is the only such four-dimensional
manifold with both the proper symmetry and the correct asymptotic
behavior.  In Sec.~7 we return to the case of an arbitrary number of
distinct fundamental monopoles.  We show that the asymptotic metric
can be smoothly continued to all values of the intermonopole
distances, and that it has all of the symmetry properties required of
the exact metric.  Hence, we conjecture that this form is, in fact,
exact.  Section 8 contains some concluding remarks.

\section{BPS Monopoles in $SU(2)$ Gauge Theory}

We begin by recalling some properties of the BPS solutions \cite{bps}
in an $SU(2)$ theory spontaneously broken to $U(1)$.  We fix the
normalization of the electric and magnetic charges $g$ and $q$ are
defined by writing the large distance behavior of the electromagnetic
field strength (in radial gauge) as
\begin{equation} 
 B_i^a = \frac{ {\hat r}_i{\hat r}_a g}{ 4\pi r^2},\quad
 E_i^a = \frac{ {\hat r}_i {\hat r}_a q}{ 4\pi r^2} .
\end{equation}
The dyon solution carrying one unit of magnetic charge (i.e.,
$g=4\pi/e$, with $e$ being the gauge coupling) may be written as
\begin{eqnarray} 
\Phi^a({\bf r}) &=& {\hat r}^a K(r; v),\nonumber \\
A_i^a ({\bf r}) &=& \epsilon_{iak} {\hat r}^k \left(A(r; v) -
   \frac{1}{ er} \right),    \nonumber \\
A_0^a ({\bf r}) &=& \frac{q}{ \sqrt{g^2 +q^2}} \Phi^a,
\label{dyonsolution}
\end{eqnarray}
where $v$ is the asymptotic magnitude of the Higgs field and 
\begin{eqnarray} 
K(r;v) &=&  v \coth(evr \eta)  - \frac{1}{ er \eta},\nonumber \\
A(r; v) &=& \frac{ v \eta}{ \sinh(evr\eta) }  ,
\end{eqnarray}
with 
\begin{equation}
\eta = \frac{g}{ \sqrt{g^2 +q^2}}.
\end{equation}
The mass of this dyon is 
\begin{equation} 
      \tilde{m} = v  \sqrt{g^2 +q^2}.
\end{equation}

This radial gauge form of the solution makes the spherical symmetry of
the dyon manifest, in that the effects of a spatial rotation can be
entirely compensated by a global $SU(2)$ gauge transformation.
However, for understanding the nature of the solutions with higher
magnetic charges, it is useful to
apply a gauge
transformation that brings the Higgs field to a constant direction in
internal space.  This introduces a Dirac string singularity that can
be chosen to lie along the $z$-axis. In cylindrical coordinates the
purely magnetic solution then takes the form
\begin{eqnarray} 
\Phi^a &=& \delta^{a3}  K(r ;v),\nonumber\\
A^3 & = &  -\frac{1}{e}\left(\frac{z}{r}\pm 1\right) 
d\phi  , \nonumber\\
W & \equiv & {1 \over \sqrt{2}}( A^1 + i A^2 ) 
      =  {ie^{\pm i\phi+i\gamma}} \frac{1}{\sqrt{2}} A(r;v)
\left(\frac{z}{r}d\rho-i\rho d\phi-
\frac{\rho}{r}dz\right ),    
\end{eqnarray} 
where the upper and lower signs correspond to solutions with Dirac
strings lying along the positive and negative $z$-axes, respectively.
Although the spherical symmetry of the solution is obscured in this
gauge, the action of the unbroken $U(1)$ is made clearer.  In
particular, a global $U(1)$ transformation simply shifts the arbitrary
phase $\gamma$, while the dyon solutions can be obtained by allowing
$\gamma$ to vary uniformly with time.  The fact that $\gamma$ has a period of
$2\pi$ implies that in the quantized theory $q$ must be an integer multiple
of $e$.  

    The solutions with higher magnetic charge can all be understood as
multimonopole solutions. To begin, consider a doubly charged solution
corresponding to two monopoles separated by a distance $R$.  If we
work in a gauge with uniform Higgs field orientation, a first
approximation to the solution can be obtained by superimposing two
single monopole solutions, provided that the two Dirac strings do not
overlap.  Hence, let us combine a solution centered about the point
$(0,0,R/2)$ with its Dirac string running upward with one 
centered at $(0,0,-R/2)$ with a Dirac string running downward
Using an obvious notation, we can write the gauge
fields of the solution as
\begin{equation} 
A_j = A_j^{(1)} +  A_j^{(2)} + \delta A_j,\quad
W_j = W_j^{(1)} +  W_j^{(2)} + \delta W_j  .
\end{equation}
Before superimposing the Higgs fields, we must first separate out the
asymptotic value $\Phi_0$.  Defining $\Delta\Phi^{(i)}({\bf r}) = 
\Phi^{(i)}({\bf r})-\Phi_0$ for the Higgs field associated with each
monopole, we have
\begin{equation} 
\Phi =  \Phi_0+ \Delta\Phi^{(1)} + \Delta\Phi^{(2)} + \delta \Phi .
\end{equation}
For $R \gg 1/ev$, we expect that $\delta A_j$, $\delta W_j$, and $\delta
\Phi$ will be small corrections that can be determined perturbatively
by substituting these expressions into the BPS field 
equations.\footnote{To completely determine these fields, 
one must both impose a
gauge condition and require that they be orthogonal to the one-monopole
zero modes about each of the component monopoles.}  

Contrary to what one might have expected, this solution is not
symmetric under rotations about the line joining the centers of of the
two monopoles.  To see this, note that whereas the $A_j^{(i)}$ and
$\Delta\Phi^{(i)}$ are explicitly independent of the azimuthal angle
$\phi$, the $W_j^{(i)}$ are of the form
\begin{eqnarray} 
 W_j^{(1)}  &=& f^{(1)}_j(\rho, z) e^{i(\phi +\gamma_1)},\nonumber\\
 W_j^{(2)} &=& f^{(2)}_j(\rho, z) e^{i(-\phi +\gamma_2)} .
\end{eqnarray}
The effect on $W_j^{(1)}$ of a rotation about the
$z$-axis can be compensated by a global $U(1)$ rotation by an equal
and opposite amount, but this just doubles the phase change of
$W_j^{(2)}$; alternatively, one can leave $W_j^{(2)}$ unchanged and
shift the phase of $W_j^{(1)}$.  The best one can do in compensating
the effects of the rotation is to apply a $z$-dependent $U(1)$
transformation that shifts the phase in one sense for $z >0$ and in
the other for $z<0$; because the magnitudes of the $W_j$ fall
exponentially at large distances from the center of the monopoles, the
violations of axial symmetry that remain after this gauge
transformation are exponentially small.  An explicit gauge-invariant
measure of the asymmetry is $\left(W_j^{(2)}({\bf r})\right)^*
W_j^{(1)}({\bf r}) \sim e^{2i\phi}$; the magnitude of this quantity is
everywhere of order $e^{-evR}$ or smaller.

      When the two monopoles are not widely separated, superposition
does not even give a good first approximation to the solution and more
sophisticated techniques are needed to obtain the solutions.  Using
such methods, one finds that the rotational asymmetry about the axis
joining the two monopoles persists as long as the centers of the two
monopoles do not coincide.  If the centers do coincide, the solution 
has a single axis of rotational symmetry; general arguments show that
spherically symmetric solutions are impossible \cite{guth}.

\section{BPS Monopoles for Larger Gauge Groups}

Recall that a basis for the Cartan subalgebra of a rank $k$ gauge
group can be chosen to be $k$ generators $H_i$ normalized so that
\begin{equation} 
{\rm tr}\, H_i H_j =\delta_{ij} .
\end{equation}
One can then define raising and lowering operators 
$E_{{\mbox{\boldmath $\alpha$}}}$ obeying 
\begin{equation}   
[{\bf H} , E_{\mbox{\boldmath $\alpha$}}] = 
{\mbox{\boldmath $\alpha$}} E_{\mbox{\boldmath $\alpha$}} ,
\end{equation}
and normalized so that 
\begin{equation}    
[E_{\mbox{\boldmath $\alpha$}} , E_{-{\mbox{\boldmath $\alpha$}}}] = 
{\mbox{\boldmath $\alpha$}} \cdot {\bf H} .
\end{equation}
The roots ${\mbox{\boldmath $\alpha$}}$ may be viewed as vectors
forming a lattice in a $k$-dimensional Euclidean space.  It is always
possible to choose a basis of $k$ simple roots for this lattice in
such a way that all other roots are linear combinations of the simple
roots with integer coefficients all of the same sign; a root is called
positive or negative according to this sign.  A set of simple roots of
particular importance is defined as follows.  Let $\Phi_0$ be the
asymptotic value of the Higgs field in some fixed direction (say, the
positive $z$-axis).  We may choose $\Phi_0$ to lie in the Cartan
subalgebra and then define a vector $\bf h$ by
\begin{equation}     
       \Phi_0 =  {\bf h}\cdot  {\bf H} .
\end{equation}
In this paper we are concerned with the case of maximal symmetry
breaking, with the gauge group spontaneously broken to $U(1)^k$.  This
is achieved if and only if $\bf h$ has a nonzero inner product with
all of the roots.  When this is so, there is a unique set of simple
roots ${{\mbox{\boldmath $\beta$}}}_a$ that satisfy the requirement
that ${\bf h} \cdot {\mbox{\boldmath $\beta$}}_a $ be positive for all
$a$; we shall use this basis for the remainder of this paper.

Asymptotically, the magnetic field must commute with the Higgs field.
Hence, in the direction chosen to define $\Phi_0$, it must be of the
form 
\begin{equation} 
B_k =  \frac{{\hat r}_ k}{ 4\pi r^2 } {\bf g}\cdot {\bf H } .
\end{equation}
Topological arguments lead to the quantization condition \cite{topology}
\begin{equation}    {\bf g} = \frac{4\pi}{e} \sum_{a=1}^k  n_a 
{{\mbox{\boldmath $\beta$}}}_a^* ,
\end{equation}
where 
\begin{equation}      
{{\mbox{\boldmath $\beta$}}_a^*} = \frac{ 
{\mbox{\boldmath $\beta$}}_a}{{{\mbox{\boldmath $\beta$}}_a^2}}  .
\end{equation} 
are the duals of the simple roots and the integers $n_a$ are the
topologically conserved charges corresponding to the homotopy class of
the scalar field at spatial infinity.

Monopole solutions carrying a single unit of topological charge can be
obtained by simple embeddings of the $SU(2)$ solution. Each simple
root ${\mbox{\boldmath $\beta$}}_a$ defines an $SU(2)$ subgroup with
generators
\begin{eqnarray} 
t^1 &=& \frac{1 }{ \sqrt{2{{\mbox{\boldmath $\beta$}}_a^2}}} 
        \left(E_{{\mbox{\boldmath $\beta$}}_a} + 
        E_{-{\mbox{\boldmath $\beta$}}_a} \right), \nonumber \\
t^2 &=& -\frac{i}{ \sqrt{2{{\mbox{\boldmath $\beta$}}_a^2}}} 
        \left(E_{{\mbox{\boldmath $\beta$}}_a} - 
        E_{-{\mbox{\boldmath $\beta$}}_a} \right), \nonumber \\
t^3 &=&  {{\mbox{\boldmath $\beta$}}_a^*}  \cdot  {\bf H} . 
\label{sub}
\end{eqnarray}
If $\Phi^s({\bf r};v)$ and $A_i^s({\bf r};v)$ ($s=1,2,3$) is
the $SU(2)$ solution corresponding to a Higgs expectation value $v$,
then 
\begin{eqnarray} 
A_i({\bf r}) &=& \sum_{s=1}^3 A_i^s({\bf r};\,{\bf h}
\cdot {\mbox{\boldmath $\beta$}}_a ) t^s ,   \nonumber\\
\Phi({\bf r})&=& \sum_{s=1}^3 \Phi^s({\bf r};\,{\bf h}
\cdot {\mbox{\boldmath $\beta$}}_a ) t^s 
 + ( {\bf h} - {\bf h}\cdot {{\mbox{\boldmath $\beta$}}_a^*} \,\,
{\mbox{\boldmath $\beta$}}_a)\cdot {\bf H} , \label{embed}
\end{eqnarray}
is a solution with topological charges
\begin{equation}    
n_b =\delta_{ab}  ,
\end{equation}
and mass
\begin{equation}   
m_a = \frac{4\pi}{e} {\bf h}\cdot {{\mbox{\boldmath $\beta$}}_a^*} .
\end{equation}
We will refer to such solutions as fundamental monopoles \cite{erick}.
As in the $SU(2)$ case, there are dyon solutions corresponding to these
fundamental monopoles, with their electric charges quantized so that 
asymptotically
\begin{equation} 
E_k =  \frac{e n{\hat r}_ k}{ 4\pi r^2 } 
 {{\mbox{\boldmath $\beta$}}_a}\cdot {\bf H }
\end{equation}
for some integer $n$.

    Solutions corresponding to several widely separated fundamental
monopoles can be constructed in a manner similar to that described
above for the $SU(2)$ case.  A notable difference, which will be of
importance later, concerns the symmetry of the two-monopole solutions.
In the $SU(2)$ case these failed to be axially symmetric because any
gauge transformation that compensated the effects of a spatial
rotation on one of the monopoles shifted the phase of the other
monopole in the wrong direction.  By contrast, if the two are
different fundamental monopoles, it is always possible to find two
gauge transformations such that each gives the required phase shift on
one of the monopoles and leaves the other invariant.  As a result, the
superposition construction yields solutions 
with exact axial symmetry \cite{ath}.
(Note that ${\rm Tr} \left(W_j^{(2)}\right)^* W_j^{(1)}$, the analogue
of the gauge-invariant measure of the asymmetry in the $SU(2)$ case,
vanishes identically.)

    A second difference from the $SU(2)$ case is the existence of
relatively simple, spherically symmetric solutions that can be
interpreted as superpositions of several fundamental monopoles at the
same point.  These are given by Eqs.~(\ref{sub}) and (\ref{embed}), 
but with a composite root ${\mbox{\boldmath $\alpha$}}$, rather than a 
simple root ${\mbox{\boldmath $\beta$}}_a$, defining the
$SU(2)$ subgroup.  The coefficients $n_a$ in the expansion
\begin{equation}   
    {\mbox{\boldmath $\alpha$}}^* =  \sum_{a=1}^k  
n_a {{\mbox{\boldmath $\beta$}}_a^*}  \label{compositeroot}
\end{equation}
are the topological charges of the solution, while the mass is
\begin{equation} 
   m = \sum_a n_a m_a .
\end{equation}
Although the mass and topological charge of these solutions are
consistent with their interpretation as superpositions of several
noninteracting monopoles, one might still ask why these spherically
symmetric solutions should be viewed on such a different basis than
those obtained from the simple roots.  An answer is obtained by
counting the normalizable zero modes about these solutions.  After
gauge fixing, the number of zero modes about an arbitrary BPS solution 
is \cite{erick}
\begin{equation} 
      N = 4 \sum_{a=1}^k n_a .
\end{equation}
Thus, each of the fundamental monopoles has four zero modes, three
corresponding to spatial translations and the fourth to a global
$U(1)$ gauge rotation.  By contrast, the solutions based on composite
roots all have additional zero modes, with the number precisely that
expected if these are in fact superpositions of several fundamental
monopoles.

The zero modes about the spherically symmetric solutions clearly must
arrange themselves in angular momentum multiplets.  For each of the
fundamental monopoles the four zero modes divide into a triplet (the
spatial translation modes) and a singlet (the global gauge rotation
mode).   Consider next an embedding solution 
corresponding to a composite root ${\mbox{\boldmath $\alpha$}}$ for which
the expansion Eq.~(\ref{compositeroot}) has only two nonvanishing $n_a$,
both equal to unity.  Of the eight zero modes about such a solution,
four keep the solution within the embedding subgroup; these are 
just the overall translation and global gauge rotation modes.  The
remaining four separate into two doublets that transform in 
opposite fashion under the ``hypercharge'' $U(1)$ generated by $( {\bf h}
- {\bf h}\cdot {{\mbox{\boldmath $\beta$}}_a^*} \,\, {\mbox{\boldmath
$\beta$}}_a)\cdot {\bf H}$.  (This implies, by the way, that none of these
modes correspond in a simple way to separation of the centers of the
constituent monopoles; such separation only occurs at second order in the
deviations from the spherically symmetric solution.)   More generally,
one can show that, for any embedded solution based on a root for which the
nonzero $n_a$ are all unity, the $N-4$ zero
modes that lie outside the embedding subgroup are all arranged in pairs
of doublets carrying opposite hypercharges.  This is not the case if any 
of the $n_a$ are greater than unity.

\section{Widely Separated Fundamental Monopoles}

The metric on the moduli space determines the motions of slowly moving
dyons. Conversely, the form of the moduli space metric can be inferred from
a knowledge the interactions between the dyons.   These become quite
complicated when several dyons approach one another.  However, for finding
the metric in the regions of the moduli space corresponding to large
intermonopole distances it is sufficient to examine the long-range pairwise
interactions between widely separated dyons.  This analysis has been carried
out previously for $SU(2)$ \cite{gary}\cite{manton}; 
in this section we will extend it to larger gauge groups. 

We begin by considering the interactions between a single pair of
$SU(2)$ dyons, with positions ${\bf x}_j$ and velocities ${\bf v}_j$ and
carrying $U(1)$ magnetic charges $g_j$ and  electric charges $q_j$
($j = 1 ,2$).  If the separation between the dyons is much larger
than the radius of a monopole core, the electromagnetic interactions between
them can be well approximated by the standard results for a pair of
moving point charges.  There is also a long-range scalar force that is
manifested as a position-dependent shift in the dyon mass.  Recall that
the mass of an isolated dyon is $v \sqrt{g^2 + q^2}$, where $v$ is the
magnitude of the vacuum expectation value of the Higgs field (in $SU(2)$ 
theory).  When there is a second dyon present, its scalar field must be 
added to $v$ in this formula.  To be more precise, Lorentz
transformation of the dyon solution of Eq.(\ref{dyonsolution}) reveals that
the  magnitude of the Higgs field about a moving dyon is 
\begin{equation}
|\Phi_0  + \Delta\Phi^{(j)}({\bf x})| = v- \frac{1}{4\pi |{\bf x}-{\bf x}_j|} 
\sqrt{1- {\bf v}_j^2} \sqrt{g_j^2 +q_j^2}+ O(r^{-2}) . \label{higgs}
\end{equation}
The effective mass of dyon 1 in the presence of dyon 2 can then be
written as 
\begin{equation}
\sqrt{g_1^2 + q_1^2}\;|\Phi_0 +  \Delta\Phi^{(2)}({\bf x}_1) | .
\end{equation}

These effects of these interactions on dyon 1 are described by the
Lagrangian  
\begin{eqnarray}
L^{(1)}_{SU(2)} &=& \sqrt{g_1^2 + q_1^2}\;|\Phi_0 +  
\Delta\Phi^{(2)}({\bf x}_1) |
 \sqrt{1- {\bf v}_1^2}\nonumber \\
    &+& q_1 \left[ {\bf v}_1 \cdot {\bf A}^{(2)}({\bf x}_1) 
    -  A_0^{(2)}({\bf x}_1)  \right]\nonumber \\
    &+& g_1 \left[ {\bf v}_1 \cdot \tilde{\bf A}^{(2)}({\bf x}_1) 
    -  \tilde A_0^{(2)}({\bf x}_1)  \right] . \label{one}
\end{eqnarray}
Here ${\bf A}^{(2)}$ and $A_0^{(2)}$ are the ordinary vector and scalar
electromagnetic potentials due to charge 2, while 
$\tilde{\bf A}^{(2)}$ and $\tilde A_0^{(2)}$ are dual potentials defined
so that ${\bf E} = - {\bf \nabla \times \tilde A}$ and ${\bf B} = -{\bf 
\nabla} \tilde A_0 - \partial \tilde{\bf A}/\partial t$.  These potentials
can be obtained by standard methods \cite{manton}.  
Substituting the results, together with
Eq.~(\ref{higgs}) for the Higgs field, into Eq.~(\ref{one}) and 
keeping only terms of up to second order in $q_j$ or ${\bf v}_j$, we obtain
\begin{eqnarray}
L^{(1)}_{SU(2)} &=& -m_1 \left( 1 - \frac{1}{2}{\bf v}_1^2 + 
\frac{q_1^2 }{2 g_1^2} \right) \nonumber\\
& -& \frac{g_1g_2 }{ 8\pi r_{12} } \left[({\bf v}_1 -{\bf v}_2)^2
- \left(\frac{q_1 }{g_1} - \frac{q_2}{ g_2}\right)^2 \right] \nonumber \\
&-&\frac{1}{ 4\pi } (g_1q_2 - g_2q_1) ({\bf v}_2 -{\bf v}_1)\cdot {\bf w}_{12},
\end{eqnarray}
where $m_1 = v g_1$, $r_{12}= |{\bf x}_1 - {\bf x}_2|$ and ${\bf
w}_{12}$ is a Dirac monopole potential, defined so that  
\begin{equation}
    {\bf \nabla}_1 \times {\bf w}_{12}({\bf x}_1 -{\bf x}_2)  = -
     \frac{ {\bf x}_1 -{\bf x}_2 }{ r_{12}^3 }.
\end{equation}

We now want to generalize this to the case of a larger group, with
dyon $j$ being obtained by embedding the $SU(2)$ solution via 
the subgroup defined by the simple root ${\mbox{\boldmath $\alpha$}}_j$. 
Working in a gauge where the Higgs field is everywhere in the Cartan
subalgebra, we use the long-range behavior of the electromagnetic field
strengths
\begin{equation}
{\bf B}^{(j)} = g_j\,  ({\mbox{\boldmath $\alpha$}}_j^*\cdot {\bf H})\,
\frac{({\bf x}-{\bf x}_j)}{4\pi \,|{\bf x}-{\bf x}_j|^3},\qquad
{\bf E}^{(j)} = q_j\,  ({\mbox{\boldmath $\alpha$}}_j^*\cdot {\bf H})\,
\frac{({\bf x}-{\bf x}_j)}{4\pi \,|{\bf x}-{\bf x}_j|^3} ,
\end{equation}
to define the magnetic and electric charges $g_j$ and $q_j$. (Note
that $q_j$ multiplies ${\mbox{\boldmath $\alpha$}}_j^*\cdot \bf H$
rather than ${\mbox{\boldmath $\alpha$}}_j\cdot \bf H$.  This turns out to
be convenient for writing the Lagrangian, but leads to the somewhat
unusual quantization condition that $q_j$ be an integer multiple of $e
{\mbox{\boldmath $\alpha$}}^2_j$.  For $SU(2)$ this agrees with our
previous convention if the single root ${\mbox{\boldmath $\alpha$}}$
is take to have unit length.) In the absence of other dyons, the
long-range Higgs field of this dyon would be
\begin{equation}
\Phi^{(j)} = {\bf h} \cdot {\bf H} -\frac{({\mbox{\boldmath
$\alpha$}}_j^*\cdot {\bf H})}{ 4\pi\, |{\bf x}-{\bf x}_j| }  \sqrt{1- {\bf
v}_j^2}\sqrt{g_j^2 +q_j^2} , 
\end{equation}
and its mass would be
\begin{equation}
{\tilde{m}}_j =  ({\mbox{\boldmath $\alpha$}}_j^*\cdot {\bf h})  
\sqrt{g_j^2 +q_j^2} . \label{mass} 
\end{equation}

The electromagnetic interactions between the two dyons can all be traced back
to terms in the fundamental field theory Lagrangian of the form ${\rm tr}
F^{(i)}_{\mu\nu} F^{(j)\mu\nu}$ and hence acquire multiplicative
factors of
\begin{equation}
{\rm tr}\; \left(({\mbox{\boldmath $\alpha$}}^*_1 \cdot {\bf H})
( {\mbox{\boldmath $\alpha$}}^*_2 \cdot {\bf H})\right)
 = {\mbox{\boldmath $\alpha$}}^*_1 \cdot {\mbox{\boldmath $\alpha$}}^*_2 .
\end{equation}
The shift in the mass of dyon $1$ due to the Higgs field of dyon $2$
is found by replacing the factor of ${\mbox{\boldmath $\alpha$}}^*_1 \cdot 
{\bf h}$ in Eq.~(\ref{mass}) by
\begin{equation}
{\mbox{\boldmath $\alpha$}}_1^* \cdot  {\bf h} - 
\frac{{\mbox{\boldmath $\alpha$}}_1^*\cdot
{\mbox{\boldmath $\alpha$}}_2^*}{ 4\pi r_{12}} 
\sqrt{1- {\bf v}_2^2} \sqrt{g_2^2 +q_2^2} .
\end{equation} 
{From} this we see that the scalar interaction terms acquire the same
factor of ${\mbox{\boldmath $\alpha$}}^*_1 \cdot 
{\mbox{\boldmath $\alpha$}}^*_2$ as the electromagnetic terms.
Aside from this factor, the long-range interactions are exactly as in
the $SU(2)$ case, and so $L^{(1)}_{SU(2)}$ can be generalized to a higher
rank group simply by replacing $1/r_{12}$ by ${\mbox{\boldmath $\alpha$}}_1^* 
\cdot{\mbox{\boldmath $\alpha$}}_2^*/r_{12}$ and ${\bf w}_{12}$ by 
${\mbox{\boldmath $\alpha$}}_1^* \cdot{\mbox{\boldmath $\alpha$}}_2^*{\bf
w}_{12}$.  

The extension to an arbitrary number of dyons is straightforward, provided
that their mutual separations are all large.  Since we are considering
fundamental dyons that all carry unit magnetic charges, we can set all of
$g_j$ equal to $g=4\pi/e$.   The Lagrangian
obtained by adding all the pairwise interactions can be written as 
\begin{equation}
L = \frac{1}{2} M_{ij} \left( {\bf v}_i \cdot {\bf v}_j 
      - {q_iq_j \over g^2}   \right)  + \frac{g }{4\pi} q_i
      {\bf W}_{ij}\cdot {\bf v}_j ,
\end{equation}
where
\begin{eqnarray}
M_{ii} &=& m_i  - \sum_{k\ne i} \frac{g^2 {\mbox{\boldmath $\alpha$}}_i^* 
\cdot {\mbox{\boldmath $\alpha$}}_k^*}{ 4\pi r_{ik}},\nonumber \\
M_{ij} &=&\frac{g^2 {\mbox{\boldmath $\alpha$}}_i^* \cdot 
{\mbox{\boldmath $\alpha$}}_j^*}{ 4\pi r_{ij}}\qquad
\hbox{\hskip 1cm if $i\neq j$},
\end{eqnarray}
with $m_i=g\,\mbox{\boldmath $\alpha$}_i^*\cdot {\bf h}$, and
\begin{eqnarray}
{\bf W}_{ii}&=&-\sum_{k\neq i}{\mbox{\boldmath $\alpha$}}_i^*\cdot
{\mbox{\boldmath $\alpha$}}_k^*{\bf w}_{ik},\nonumber\\
{\bf W}_{ij}&=&{\mbox{\boldmath $\alpha$}}_i^*\cdot
{\mbox{\boldmath $\alpha$}}_j^*{\bf w}_{ij}\qquad
\hbox{\hskip 1cm if $i\neq j$}.
\end{eqnarray}
with ${\bf w}_{ij}$ being value at ${\bf x}_i$ of the Dirac potential
due to the $j$th monopole. The $q$-independent part of the monopole
rest energies has been omitted.

     To obtain the moduli space metric, we need a Lagrangian that is
purely kinetic; i.e., one in which all terms are quadratic in velocities. 
This can be done by interpreting the $q_j/e$ as conserved momenta conjugate
to cyclic angular variables $\xi_j$; because  
$q_j/e$ is quantized in  integer multiples of ${\mbox{\boldmath
$\alpha$}}_j^2$, the period of $\xi_j$ must be $2\pi/{\mbox{\boldmath
$\alpha$}}_j^2$. Thus, if we make the identification
\begin{equation}
q_i/e = \frac{g^4}{(4\pi)^2}(M^{-1})_{ij}(\dot\xi_j + 
{\bf W}_{jk}\cdot {\bf v}_k ),
\end{equation}
the desired Lagrangian ${\cal L}$ is the Legendre
transform  
\begin{eqnarray}
{\cal L}&=& L + \sum_j \dot{\xi}_jq_j/e \nonumber \\
 &=& \frac{1}{2} M_{ij}  {\bf v}_i \cdot {\bf v}_j +  
    \frac{g^4}{2(4\pi)^2}(M^{-1})_{ij} 
    \left( \dot\xi_i + {\bf W}_{ik}\cdot {\bf v}_k \right)
    \left( \dot\xi_j + {\bf W}_{jl}\cdot {\bf v}_l \right). \label{L}
\end{eqnarray} 
{From} this we immediately obtain the large separation 
approximation to the moduli space metric,
\begin{equation}
{\cal G}=\frac{1}{2}M_{ij}d{\bf x}_i\cdot d{\bf x}_j+\frac{g^4}{2(4\pi)^2} 
(M^{-1})_{ij}(d\xi_i+{\bf W}_{ik}\cdot d{\bf x}_k)(d\xi_j+{\bf W}_{jl}
\cdot d{\bf x}_l) .\label{metric}
\end{equation}
Note that this metric is equipped with a number of $U(1)$ isometries, each
of which is generated by the constant shift of one of the $\xi_j$'s.

     For the case of $SU(2)$, one can easily see that this approximation 
to the moduli space metric cannot be exact \cite{manton2}.  First of
all, it develops singularities if any of the intermonopole distances
becomes too small, whereas the moduli space metric should be
nonsingular.  Second, for the case of two monopoles the approximate
metric is independent of the relative phase angle $\xi_1
-\xi_2$.  If this isometry were exact, the two-monopole solutions
would be axially symmetric, which we know is not the case.  

     Neither of these objections arises for the moduli space
corresponding to a collection of several distinct fundamental
monopoles in a larger group, provided that each corresponds to a
different simple root.  In Secs.~5 and 7 we will show that the metric
is nonsingular for all values of the $r_{ij}$ in such cases.  Further,
we argued in the previous section that solutions with two different
fundamental monopoles are axially symmetric, and so the $U(1)$
isometries of the metric $\cal G$ are just what we want.

In the next two sections we will concentrate on the two-monopole case and
show that the metric obtained in this section must be the exact moduli space
metric.  We will then return to the case of three or more distinct
fundamental monopoles and argue that $\cal G$ is likely to be exact for this
case also.

\section{A Pair of Distinct Fundamental Monopoles}

We now specialize the results of the previous section to the case of two
fundamental monopoles.  There are three possibilities, distinguished by
the sign of $\lambda= -2 {\mbox{\boldmath $\alpha$}}_1^*\cdot
{\mbox{\boldmath $\alpha$}}_2^*$. This is negative only if both monopoles 
are based on the same simple root. In this case the moduli space is the 
same as that for two $SU(2)$ monopoles \cite{atiyah},
where we know that the asymptotic metric develops a singularity at
small monopole separations.  For two distinct monopoles, $\lambda=0$ if
their roots are not connected in the Dynkin diagram of the original gauge
algebra and $\lambda >0$ if the roots are so connected.  In the former case
there are clearly no interactions between the monopoles, and the moduli
space is simply the product of two one-monopole moduli spaces.  This leaves
only the case $\lambda>0$, to which we now turn.  We then have two
distinct roots ${\mbox{\boldmath $\alpha$}}_1$ and ${\mbox{\boldmath
$\alpha$}}_2$.  Because these are both simple, the
combinations $\lambda{\mbox{\boldmath $\alpha$}}_1^2=-2
{\mbox{\boldmath $\alpha$}}_1\cdot{\mbox{\boldmath $\alpha$}}_2/
{\mbox{\boldmath $\alpha$}}_2^2$ and $\lambda{\mbox{\boldmath $\alpha$}}_2^2
=-2{\mbox{\boldmath $\alpha$}}_1\cdot{\mbox{\boldmath $\alpha$}}_2/
{\mbox{\boldmath $\alpha$}}_1^2$ are positive integers no larger than $3$.
On the other hand, $(\lambda{\mbox{\boldmath $\alpha$}}_1^2)(\lambda
{\mbox{\boldmath $\alpha$}}_2^2)/4=({\mbox{\boldmath $\alpha$}}_1\cdot
{\mbox{\boldmath $\alpha$}}_2)^2/{\mbox{\boldmath $\alpha$}}_1^2
{\mbox{\boldmath $\alpha$}}_2^2$ must be less than one. Thus, without loss of 
generality,  we can set $\lambda{\mbox{\boldmath $\alpha$}}_1^2
=p$ and $\lambda {\mbox{\boldmath $\alpha$}}_2^2=1$, where $p$ is
either 1, 2, or 3.  
In particular, the rank-two gauge groups $SU(3)$, $SO(5)$, and $G_2$
yield $p=1$, 2, and 3, respectively.

We begin by separating center-of-mass from  
relative variables.  For the spatial coordinates we introduce the usual
quantities  \begin{equation}
 {\bf R} = \frac{m_1 {\bf x}_1 + m_2 {\bf x}_2}{m_1+m_2},\qquad
 {\bf r} = {\bf x}_1-{\bf x}_2 .
\end{equation}
To separate the phase variables, we first introduce the total charge $q_\chi$
and the relative charge  $q_\psi$ as linear combinations of the electric
charges $q_1$ and $q_2$,  
\begin{equation} 
q_\chi = \frac{(m_1 q_1 + m_2 q_2)}{e\,(m_1+m_2)},\qquad
q_\psi =  \frac{\lambda(q_1-q_2)}{2e}.
\end{equation}
(We noted in the previous section that $q_j$ is quantized in integer
units of $e {\mbox{\boldmath $\alpha$}}_j^2$.  It follows that the
relative charge $q_\psi$ has eigenvalues $0, \pm 1/2, \pm 1,...$.  In
contrast, the total charge $q_\chi$ is not quantized unless the ratio
$m_1/m_2$ is a rational number.)
The conjugate variables to these charges are 
\begin{equation}
\chi =  (\xi_1 + \xi_2),\qquad
\psi = \frac{2(m_2\xi_1 - m_1\xi_2)}{\lambda \,(m_1+m_2)}.
\end{equation}

      When expressed in these variables, the Lagrangian separates into
a sum of two terms,
\begin{equation}
{\cal L}^{(2)} = {\cal L}_{{\rm cm}}^{(2)} + {\cal L}_{{\rm rel}}^{(2)},
\end{equation}
where
\begin{equation} 
{\cal L}_{{\rm cm}}^{(2)} = \frac{1}{2}(m_1+m_2)\dot{{\bf R}}^2 + 
\frac{g^4 }{32\pi^2 (m_1+m_2) } \dot{\chi}^2 ,
\end{equation}
and ${\cal L}_{{\rm rel}}^{(2)}$ is the rest:
\begin{equation} 
{\cal L}_{{\rm rel}}^{(2)} = \frac{1}{2} \biggl( \mu + 
\frac{g^2\lambda}{8\pi r}\biggr)\,\dot{{\bf r}}^2
+ \frac{1}{2} \left(\frac{g^2\lambda}{8\pi}\right)^2 
\biggl( \mu+ \frac{g^2\lambda}{8\pi r}\biggr)^{-1} 
(\dot{\psi} + {\bf w}({\bf r})\cdot \dot{{\bf r}}) ^2 .
\end{equation}
Here $\mu$ is the reduced mass and ${\bf w}({\bf r})={\bf w}_{12}({\bf r})$. 

The center-of-mass part of the moduli space 
is parameterized by $\bf R$ and $\chi$ with the metric,
\begin{equation}
{\cal G}_{\rm cm}^{(2)}=\frac{m_1+m_2}{2}\left(d{{\bf R}}^2 + \frac{g^4
}{16\pi^2 (m_1+m_2)^2 }d{\chi}^2\right),
\end{equation}
and is a flat four-dimensional manifold.  Defining $r_0\equiv g^2\lambda/8\pi
\mu $, we find that the dynamics of ${\cal L}_{\rm rel}^{(2)}$ is reproduced
by the geodesic motion on a  four-dimensional manifold with metric  
\begin{equation}
{\cal G}_{\rm rel}^{(2)} = \frac{\mu}{2}\left( (1+ r_0/r)\,d{{\bf r}}^2 
+ r_0^2(1+ r_0/r)^{-1}(d{\psi}+ {\bf w}({\bf r})\cdot d{{\bf r}})^2 \right).
\label{gtwo}
\end{equation}
Apart from an overall rescaling, this is the metric  of a Taub-NUT manifold
with length  parameter $l=r_0/2$.   This is smooth everywhere but at $r=0$,
where there is a singularity unless $\psi$ has a period of $4\pi$.

Locally, the full moduli space is a product of these two manifolds.  
However, the periodicity of the $\xi_i$'s imposes certain
identifications on $\psi$ and $\chi$, so that the total moduli space is
obtained only after a division by a discrete group, as we now show.  In the
previous section we noted that the quantization of the charge operators
$q_i$ implied that the $\xi_i$ had periods $2\pi/{\mbox{\boldmath
$\alpha$}}_i^2$.  A shift of  $\xi_1$ by $2\pi/{\mbox{\boldmath
$\alpha$}}_1^2 $ then implies the identification  
\begin{equation} 
(\chi, \psi) = (\chi +  \frac{2\pi}{{\mbox{\boldmath $\alpha$}}_1^2}, 
 \psi + \frac{4\pi m_2}{\lambda{\mbox{\boldmath $\alpha$}}_1^2 (m_1+m_2)} ), 
\label{shift}
\end{equation}
while a $-2\pi/{\mbox{\boldmath $\alpha$}}_2^2$ shift of $\xi_2$ gives
\begin{equation} 
(\chi, \psi) =  ( \chi - \frac{2\pi}{ {\mbox{\boldmath $\alpha$}}_2^2},  
\psi + \frac{4\pi m_1}{\lambda {\mbox{\boldmath $\alpha$}}_2^2 (m_1+m_2)} ) .
\end{equation}
Then, after $p = \lambda{\mbox{\boldmath $\alpha$}}_1^2$ steps of the
first shift and one step of the second, we discover that
\begin{equation}
(\chi,\psi) = (\chi,\psi +4\pi).
\end{equation}
This shows that the $\psi$ has a period of $4\pi$, in agreement with
the half-integer quantization of $q_\psi$ noted above, and hence that
the relative coordinate metric ${\cal G}_{\rm rel}^{(2)}$ is smooth
everywhere.

Without these identifications, the surface spanned by the coordinates
$(\chi,\psi)$ (at any $r\neq 0$) would be a cylinder. However, the
identification (\ref{shift}) tells us to cut out a section of this
cylinder of length $\delta\chi=2\pi\lambda /p$ and to identify the two
rims with a twist $\delta \psi =4\pi m_2/p(m_1+m_2)$.  The result is
that the moduli space is of the form \cite{klee}
\begin{equation}
{\cal M } = R^3 \times \frac{R^1 \times {\cal M}_0}{Z} ,  \label{Roneform}
\end{equation}
where ${\cal M}_0$ is the Taub-NUT manifold with metric (\ref{gtwo}).
If the ratio of the two monopole masses is rational, $\chi$ is in fact
periodic.  For instance, when the masses are equal, the generator of
the identification map becomes $(\chi,\psi)= (\chi +2\pi\lambda /p,
\psi + 2\pi/p)$, which collapses the $R^1$ to a circle of length $4\pi
\lambda$. Accordingly, the moduli space can be written as
\begin{equation}
{\cal M} = R^3 \times \frac{S^1 \times {\cal M}_0}{Z_{2p}}.
\end{equation}
Such compactification of the $R^1$ to $S^1$ corresponds to the fact
that $p_\chi$ is quantized when the mass ratio is rational.  

The Lagrangian ${\cal L}_{\rm rel}^{(2)}$ possess a rotational
invariance.  Although the term involving $\dot\psi$ is not manifestly
invariant under rotation of the vector $\bf r$, the situation is
similar to that of a charged particle in the presence of a point-like
monopole and the effect of this last term is to simply modify the
conserved angular momentum in a familiar way \cite{manton2}:
\begin{equation}
 {\bf J}=   \biggl( \mu + \frac{g^2\lambda}{8\pi r}\biggr)
  {\bf r}\times\dot{\bf r} +q_\psi\hat{\bf r}.
\end{equation}
Note that for $q_\psi=\pm 1/2, \pm 3/2,\dots$, this angular momentum
is quantized at half-integer values, so that the rotation group is
actually $SU(2)$ instead of the naive $SO(3)$. 

{From} a more geometrical point of view, the conserved angular momentum
$\bf J$ implies a set of Killing vector fields on ${\cal M}_0$ that
generate an $SU(2)$ translational symmetry. In other words, the
spatial rotations of the monopoles induce an $SU(2)$ isometry of the
moduli space itself. Since a rotation leaves the relative distance $r$
between the two monopoles invariant, the orbit is spanned by angular
coordinates only. Furthermore, because of the extra term proportional
to $q_\psi$, the angular momentum ${\bf J}$ shifts $\psi$ in addition
to rotating the two-sphere $|{\bf r}|=r$, so the orbit is
generically three-dimensional. In fact, the three-dimensional orbit at
$r>0$ is simply an $S^3$ parameterized by the three Euler angles
$\psi$, $\theta$, and $\phi$, with the latter two spanning the two-sphere
$|{\bf r}|=r$. The only exception is at origin, $r=0$, where the orbit
collapses to a point.

We have shown in this section that for a pair of distinct fundamental
monopoles the asymptotic metric found in Sec.~4 remains nonsingular
for all values of the monopole separation.  While this means that the
asymptotic form could, in fact, be exact, it certainly not prove that
it is.  In the next section we will use the methods that Atiyah and
Hitchin used for the $SU(2)$ case to demonstrate that Eq.~(\ref{Roneform})
does indeed give the exact two-monopole moduli space metric.

\section{The Two-Monopole Moduli Space} 

Symmetry considerations tell us a good deal about the form of the
two-monopole moduli space $\cal M$.  First of all, there must be three
flat directions corresponding to overall spatial translations of the
two-monopole system.  Next, note that the dyon solutions obtained from
a time-dependent phase rotation generated by $\Phi_0$ are BPS
solutions and therefore do not interact when at rest.  This implies
the existence of a fourth flat direction, corresponding to the
coordinate $\chi$ introduced in the previous section.  Allowing for
the possibility of identifications of the sort that we found in
Sec.~5, we conclude that the space must be of the form
\begin{equation}
{\cal M}=R^3\times \frac{R^1\times {\cal M}_0}{\cal D}.
\end{equation}
where ${\cal D}$ is a discrete normal subgroup of the isometry group
of $R^1\times{\cal M}_0$.

The isometry group of ${\cal M}_0$ is also easily determined.  Since a
spatial rotation of a BPS solution about any fixed point is again a
solution, ${\cal M}_0$ must possess a three-dimensional rotational
isometry.  One linear combination of the two unbroken $U(1)$ gauge
degrees of freedom generates the translational symmetry, alluded to
above, along the overall $R^1$.  The remaining generator must then
induce a $U(1)$ isometry acting on ${\cal M}_0$.  Hence, ${\cal M}_0$
must be a four-dimensional manifold with at least four Killing vector
fields that span an $su(2)\times u(1)$ algebra.
Furthermore, we saw in the previous section that the orbits of the
rotational isometry on the asymptotic metric were three-dimensional;
clearly the exact metric must also possess this property at large $r$.

In addition, the moduli space must be hyperk\"ahler.\footnote{A 
$4n$-dimensional manifold with a metric is hyperk\"ahler if it 
possesses three covariantly constant complex structures ${\cal J}^{(k)}$
that also form a quarternionic structure, and if the metric is 
pointwise Hermitian with respect to each ${\cal J}^{(k)}$. If we recast 
the zero modes $(\delta \Phi, \delta A_i)$ into  a spinor $\Psi=\delta 
\Phi+i\tau_i\delta A_i$ where $\tau_i$'s are the Pauli matrices, the
zero mode equation is manifestly invariant under right multiplications
by the $i\tau_k$'s \cite{erick}, and this induces the almost quarternionic
structure on the moduli space. Detailed arguments that the moduli space
is hyperk\"ahler can be found in Refs.~\cite{atiyah} and \cite{jg}.}
In four dimensions this implies that the manifold must be a self-dual
Einstein manifold.  From this, together with the rotational symmetry
properties of the manifold, it follows that the metric can be written
as \cite{gibbons}
\begin{equation}
ds^2=f(r)^2\,dr^2+a(r)^2\sigma_1^2+b(r)^2\sigma_2^2+c(r)^2\sigma_3^2,
\end{equation}
where the metric functions obey 
\begin{equation}
\frac{2bc}{f}\frac{da}{dr}=b^2+c^2-a^2-2\lambda bc,\quad \hbox{and cyclic
permutations thereof},\label{abc}
\end{equation}
with $\lambda$ either 0 or 1,
while the three one-forms $\sigma_k$ satisfy
\begin{equation}
d\sigma_i=\frac{1}{2}\epsilon_{ijk}\sigma_j\wedge \sigma_k. 
\end{equation}
An explicit representation for these one-forms is 
\begin{eqnarray}
\sigma_1 &=& -\sin\psi d\theta +\cos\psi\sin\theta d\phi, \nonumber \\
\sigma_2 &=& \cos\psi d\theta +\sin\psi\sin\theta d\phi, \nonumber \\
\sigma_3 &=&  d\psi+\cos\theta d\phi.
\end{eqnarray}
The ranges of $\theta$ and $\phi$ are $[0,\pi]$ and $[0,2\pi]$,
respectively.  In order that the metric tend to the ${\cal G}_{\rm
rel}$ found in Sec.~5, the range of $\psi$ must be $[0,4\pi]$.

Following Atiyah and Hitchin \cite{atiyah}, we may reduce Eq.~(\ref{abc}) to a
two-dimensional flow equation for $x=a/b$ and $y=c/b$ in terms of
$\eta =\int fb/ac\,dr$.
\begin{equation}
\frac{dx}{d\eta}=x(1-x)(1+x-\lambda y),\qquad
\frac{dy}{d\eta}=y(1-y)(1+y-\lambda x). \label{xy}
\end{equation}
The case of $\lambda=1$ was studied in detail by Atiyah and Hitchin, where
they found only three possibilities (up to irrelevant
permutations of $\sigma_k$'s) that corresponded to complete manifolds:

\vskip 5mm
\noindent
1) $a=b=c$ case: flat $R^4$.

\noindent
2) $a=b\neq c$ case: the Taub-NUT geometry with an $SU(2)$ rotational
isometry.

\noindent
3) $a\neq b\neq c$ case: the Atiyah-Hitchin geometry with an $SO(3)$
rotational isometry.
\vskip 5mm

\noindent
The case of $\lambda=0$ leads to only one new possibility:\footnote{
In fact, as pointed out by Atiyah and Hitchin, the case of $\lambda=0$
need not be considered. When $\lambda=0$, the rotational isometry
leaves the three complex structures of the moduli space individually
invariant, but this cannot be the case for the real moduli space.}

\vskip 5mm
\noindent
4) the Eguchi-Hanson gravitational instanton \cite{eguchi}.

\vskip 5mm
\noindent
The moduli space is clearly curved, since otherwise there would be no 
interaction, and so we can exclude $R^4$. Of the remaining three, the only
one that asymptotes to the geometry of ${\cal G}_{\rm rel}^{(2)}$ is the 
Taub-NUT geometry, and thus ${\cal M}_0$ must be a Taub-NUT
manifold~[10-12]. 
With a suitable choice of $r$, its metric can be written as
\begin{eqnarray}
ds^2
&=&\left(1+\frac{2l}{r}\right)\left(dr^2+r^2\sigma_1^2+r^2\sigma_2^2\right)+
\left(\frac{4l^2}{1+2l/r}\right)\sigma_3^2\nonumber \\
&=&\left(1+\frac{2l}{r}\right)\,d{\bf r}^2+\left(\frac{4l^2}{1+2l/r}\right)
(d\psi+ \cos\theta\,d\phi)^2 .
\end{eqnarray}
(Note that $\psi$ dependence drops out of the metric coefficients so
that we have an additional Killing vector field,
$\xi_3=\partial/\partial\psi$, that generates the extra $U(1)$
isometry anticipated above.)  A comparison of this metric with the
asymptotic form ${\cal G}_{\rm rel}^{(2)}$ in Eq.~(\ref{gtwo}) reveals
that they are identical, up to a trivial rescaling and a gauge choice
for the vector potential ${\bf w}$, if we identify the length
parameter $l$ with $r_0/2=-g^2 {\mbox{\boldmath
$\alpha$}}_1^*\cdot{\mbox{\boldmath $\alpha$}}_2^*/8\pi\mu$. 

A further check that ${\cal M}_0$ must be Taub-NUT follows from the
existence of the spherically symmetric BPS solution obtained, as
described below Eq.~(\ref{compositeroot}) by embedding the $SU(2)$
monopole using the subgroup defined by the composite root whose dual
is the sum of the duals of the two relevant simple roots.  This
solution obviously corresponds to a point of ${\cal M}_0$ that is
invariant under the rotational isometry.  While there are no such
points in the Atiyah-Hitchin and Eguchi-Hanson geometries, the point
$r=0$ of the Taub-NUT metric is just what we want.  The apparent
singularity in the metric at this point turns out to be a simple
coordinate singularity, also known as a ``NUT'' singularity, that is
easily removed by introducing a new radial coordinate with
$\rho^2\simeq 8lr$ so that  
\begin{equation}
ds^2= \left[ 1 + O(\rho^2) \right]
d\rho^2+\frac{\rho^2}{4}\left(\sigma_1^2+\sigma_2^2+\sigma_3^3\right) +
O(\rho^4). 
\end{equation} 
To the leading order, this is the metric of the flat $R^4$  Euclidean
space and thus obviously smooth at the origin. (The unfamiliar factor of
$1/4$ is due to the normalization of the $\sigma_k$.)   From this
expression it is clear that the infinitesimal motions away from the
origin (i.e., the zero modes about the spherically symmetric solution)
transform under the vector representation of the $SO(4)$ symmetry of
$R^4$. In terms of the three-dimensional rotations of physical space,
this vector decomposes into a pair of doublets, in agreement with the
zero mode multiplet structure noted at the end of Sec.~3.

\section{The Smooth Moduli Space of $n$ Fundamental Monopoles}

We saw in Sec.~5 that for a pair of distinct fundamental monopoles the
asymptotic form of the moduli space metric remained nonsingular for all
values of the monopole separation, suggesting that this asymptotic
form might be the exact metric, a conjecture that was verified by the
analysis of Sec.~6.  We now return to the more general case of an
arbitrary number of distinct fundamental monopoles.  We will argue
the asymptotic form is likely to be exact in this case also.

Thus, suppose we have $n$ distinct fundamental monopoles of charges
${\mbox{\boldmath $\alpha$}}_i^*$. These $n$ ${\mbox{\boldmath
$\alpha$}}_i$'s form a subset of the simple roots of the Lie algebra
and define a subdiagram of its Dynkin diagram.  In general, this
subdiagram would be composed of several connected components. However,
monopoles that belong to one connected component have no interaction
with those belonging to others, so that the total moduli space would
be a product of moduli spaces for each connected component.  Hence, it
is sufficient to consider the case where the diagram spanned by the
${\mbox{\boldmath $\alpha$}}_i$'s is connected, and is therefore the
full Dynkin diagram of some (possibly smaller) simple gauge group, 

The moduli space metric for this $n$-monopole system must be both
hyperk\"ahler and nonsingular.  The demonstration that the asymptotic
metric is hyperk\"ahler is essentially the same as for the $SU(2)$
case treated by Gibbons and Manton \cite{gary}.  The key ingredient is that the
fact that the gradient of the scalar potential $1/r_{ij}$ is equal to
the curl of the vector potential ${\bf w}_{ij}$ for each pair of
monopoles; since the only modification we have introduced is to
multiply both potentials by ${\mbox{\boldmath $\alpha$}}_i^*\cdot
{\mbox{\boldmath $\alpha$}}_j^*$, the hyperk\"ahler property is
preserved.

We next turn to the issue of singularities.  The most obvious set of
potential singularities are the places where one or more of the $r_{ij}$
vanish; we will leave these for last.  A second possibility is that a
metric component might diverge, even though all of the $r_{ij}$ are
nonzero, because the $k\times k$ matrix $M_{ij}$ failed to be
invertible.  The final possibility is that is that the metric might
become degenerate, which would be seen as a zero of its determinant
\begin{equation}
{\rm Det}\;{\cal G} =  \left(\frac{g^2}{16\pi}\right)^{2n}
\left({\rm Det}\;M\right)^{2} .
\end{equation}
To rule out these last two possibilities it suffices to show that
${\rm Det}\, M$ is nonzero whenever the $r_{ij}$ are nonzero.  

To show this, all we need is the fact that $M_{ij}$ can be written in
the form
\begin{eqnarray}
M_{ii} &=& m_i  + \sum_{j\ne i} c_{ij} ,\nonumber \\
M_{ij} &=& -c_{ij} \qquad
\hbox{\hskip 1cm if $i\neq j$},
\end{eqnarray}
where the $c_{ij}$ are all nonnegative and the $m_i$ are all positive
definite.  For $n=2$ it is trivial to see that ${\rm Det}\, M >0$.
For $n>2$ we proceed inductively.  First, we note that the determinant
is equal to zero if all of the $m_i$ vanish.  Second, the the partial
derivative of ${\rm Det}\, M$ with respect to any one of the masses,
say $m_1$, is the determinant of the symmetric $(n-1)\times(n-1)$
matrix obtained by eliminating the first row and column. This new matrix is of
the same type as $M$ (except that now we have $m_j \rightarrow m_j + 
c_{1j} > 0$), and hence has a positive determinant by the induction
hypothesis. It then immediately follows that ${\rm Det}\, M >0$ and that
the metric is nonsingular if none of the $r_{ij}$ vanish.

Before proceeding further,
it is convenient to separate out the center-of-mass coordinates.
A crucial point to observe here is
that there are exactly $n-1$ number of links among the 
${\mbox{\boldmath $\alpha$}}_i$'s in the connected Dynkin subdiagram 
spanned by the ${\mbox{\boldmath $\alpha$}}_i$'s. Because of this, 
there are exactly $n-1$ number of unordered pair of distinct indices 
$\{i,j\}$ such that ${\mbox{\boldmath $\alpha$}}_i \cdot 
{\mbox{\boldmath $\alpha$}}_j\neq 0$.
It is easy to see that the metric coefficients
are then functions only of these ${\bf x}_i-{\bf x}_j$.  Thus introducing a 
label $A$ for these unordered pairs $\{i,j\}$, we can introduce
coordinates 
\begin{equation}
{\bf R}=\frac{\sum m_i {\bf x}_i}{\sum m_i},\qquad
{\bf r}_A={\bf x}_i-{\bf x}_j . \label{split}
\end{equation}
Similarly, we can split the electric charges $q_i$ of the $n$ monopoles into
a total charge $q_\chi$ and relative charges $q_A$, defined as
\begin{equation}
q_\chi= \frac{\sum m_i q_i}{e\sum m_i},\qquad 
q_A=\frac{\lambda_A}{2e}(q_i-q_j),
\end{equation}
with $\lambda_A\equiv -2{\mbox{\boldmath
$\alpha$}}_i^*\cdot{\mbox{\boldmath $\alpha$}}_j^*$. As in the
two-monopole case the relative charges $q_A$'s are quantized at
half-integer values, and so their conjugate variables $\psi_A$ all
have period $4\pi$.  The center-of-mass degrees of freedom $\bf R$ and
$\chi$ now decouple from the rest, and we arrive at the relative
moduli space metric,
\begin{eqnarray}
{\cal G}_{\rm rel}
&=&\frac{1}{2}\,C_{AB}\,d{\bf r}_A\cdot d{\bf r}_B\nonumber \\
&+&\frac{g^4\lambda_A\lambda_B}{2(8\pi)^2}\,(C^{-1})_{AB}\,
(d\psi_A+{\bf w}({\bf r}_A)\cdot d{\bf r}_A)
(d\psi_B+{\bf w}({\bf r}_B)\cdot d{\bf r}_B).
\end{eqnarray}
Here the $(n-1)\times (n-1)$ matrix $C_{AB}$ is 
\begin{equation}
C_{AB}=\mu_{AB}+\delta_{AB}\,\frac{g^2\lambda_A}{8\pi r_A},
\end{equation}
where $r_A=|{\bf r}_A|$ and $\mu_{AB}$ is a constant matrix that 
can be found by using
the redefinition (\ref{split}) and the free part of the Lagrangian $\cal
L$. (Note that $\mu_{AB}$ is diagonal only if $n=2$.)

This expression for ${\cal G}_{\rm rel}$ is manifestly invariant under
independent constant shifts of the periodic coordinate $\psi_A$.   The
$\partial/\partial \psi_A$'s are thus Killing vectors that generate $n-1$
$U(1)$ isometries of the relative moduli space.  These, together with the
isometry under uniform translation of the global phase $\chi$, correspond
to the action of the $k$ independent global $U(1)$ gauge rotations
generated by the ${\mbox{\boldmath $\alpha$}}_i\cdot {\bf H}$. 

We have assumed that the  ${\mbox{\boldmath $\alpha$}}_i$'s are connected
and distinct.  It is easy to see that the sum  $\sum {\mbox{\boldmath
$\alpha$}}_i^*$ is then equal to $\mbox{\boldmath $\gamma$}^*$ where
$\mbox{\boldmath $\gamma$}$ is some positive root of the group $G$.  
Embedding of the $SU(2)$ BPS monopole using the subgroup generated by
$\mbox{\boldmath $\gamma$}$ gives a spherically symmetric solution on
which $n-1$ of the $U(1)$ gauge rotations (i.e., those orthogonal to
$\mbox{\boldmath $\gamma$}\cdot {\bf H}$) act trivially.  The existence
of this solution implies that there must be a maximally 
symmetric point on the relative moduli space that is a fixed point 
both under overall rotation of the $n$-monopoles and
under the $n-1$ $U(1)$ translations.   This fixed point is just the
origin, ${\bf r}_A=0$ for all $A$.  

In the neighborhood of the origin, the factors of $1/r_A$ are all
sufficiently large that the matrix $C_{AB}$ is effectively diagonal. The
behavior of the metric ${\cal G}_{\rm rel}$ is then
\begin{equation}
{\cal G}_{\rm rel}\simeq \frac{g^2}{16\pi}\sum_A \lambda_A\left(\frac{1}{r_A}
d{\bf r}_A^2+r_A\,(d\psi_A+{\bf w}({\bf r}_A)\cdot d{\bf r}_A)^2\right),
\end{equation}
with the leading corrections linear in the $r_A$'s.  Comparing this with
the results of the two previous sections, we see that the manifold is
nonsingular at the origin, and that the four-dimensional
component corresponding to each $A$ gives the metric of $R^4$, with the 
invariant distance from the origin measured,
up to a common multiplicative constant, by $\lambda_A r_A$.  As with the
two-monopole case, this is consistent with the fact that the zero modes
about the spherically symmetric solution include $n-1$ pairs of
rotational doublets in addition to the four modes corresponding to overall
translation or gauge rotation.

Finally, we consider the points where only some of the $r_A$'s vanish; 
we distinguish those that vanish by replacing the subscript $A$ by $V$. In
inverting $C_{AB}$ to leading order, it suffices to remove all components of
$\mu_{AB}$ in the rows or the columns labeled by the $V$'s.  The
matrix $C$ is then effectively block-diagonal, and consists of the
diagonal entries $g^2\lambda_V/8\pi r_V$'s as well as a chunk of
smaller square matrices. Looking for the part of metric along the
${\bf r}_V$ and $\psi_V$ directions, we find
\begin{equation}
{\cal G}_{\rm rel}\simeq \frac{g^2}{16\pi}\sum_V \lambda_V\left(\frac{1}{r_V}
d{\bf r}_V^2+r_V\,(d\psi_V+{\bf w}({\bf r}_V)\cdot d{\bf r}_V)^2\right)
+\cdots .
\end{equation}
The terms shown explicitly indicated that the four-dimensional part
corresponding to each of the $V$'s is again a smooth $R^4$.  The
remaining terms, indicated by the ellipsis, consist of harmless
finite parts quadratic the other $d{\bf r}_A$'s and $d\psi_A$'s as
well as mixed terms that involve one of the $d{\bf r}_V$'s or
$d\psi_V$'s multiplied by a $d{\bf r}_A$ or $d\psi_A$.  The
off-diagonal metric coefficients corresponding to the latter vanish
linearly in terms of local Cartesian coordinates at ${\bf r}_V=0$, and
hence cannot introduce any singular behavior at ${\bf r}_V=0$. We
conclude that the metric ${\cal G}_{\rm rel}$, and thus ${\cal G}$, remain
smooth as any number of monopoles come close together.

\section{Discussion}

In this paper we have shown that the metric obtained from the dynamics
of well-separated fundamental monopoles is in fact smooth for all
values of the intermonopole distances, provided that the monopoles are
all based on different simple roots of the maximally broken gauge
group.  Furthermore, this metric is hyperk\"ahler, and displays the
isometries and the behavior near its rotationally symmetric fixed
point that the exact metric must have.  For the case of any two
distinct monopoles we have shown that this asymptotic metric is indeed
exact, and it seems likely that this is the case for any number of
such objects.

The asymptotic form for the metric
can be exact only if the fundamental monopoles are all distinct, since it
develops a curvature singularity if a pair of identical monopoles are
brought too close together.  However, there are some cases involving two or
more identical monopoles where the existence of spherically symmetric
solutions suggests that there may be some
simplifications compared to the general case.  While these can arise for
any group other than $SU(n)$, the simplest example occurs in 
maximally broken $SO(5)$.  If $\mbox{\boldmath $\alpha$}$ and
$\mbox{\boldmath $\beta$}$  are the long and the short simple roots for
this case,  then $2\mbox{\boldmath $\alpha$}^*+\mbox{\boldmath
$\beta$}^* = (\mbox{\boldmath $\alpha$}+\mbox{\boldmath
$\beta$})^*$, which means that there is a spherically symmetric embedded
solution with triple magnetic charge $2\mbox{\boldmath
$\alpha$}^*+\mbox{\boldmath $\beta$}^*$.  This solution must correspond to
a rotationally invariant fixed point of the eight-dimensional
relative moduli space.  If one of the $\mbox{\boldmath $\alpha$}^*$
monopoles is removed to a large distance while the others are kept
close together, the four-dimensional subspace spanned by the 
relative coordinates of the latter pair should resemble the smooth Taub-NUT
space, but if the two $\mbox{\boldmath $\alpha$}^*$ monopoles are kept
together while the $\mbox{\boldmath $\beta$}^*$ monopole is taken far away,
an approximation to the Atiyah-Hitchin geometry should emerge.  

A second open issue concerns theories for which the unbroken gauge
group has a non-Abelian factor.  The zero mode counting has been
worked out in detail for the cases where the magnetic charge has only
Abelian components.  As one might expect, the dimension of the moduli space 
is typically greater than $4 \sum \tilde n_a$, where the $\tilde n_a$ are 
the topologically conserved magnetic charges \cite{erick2}. At the
same time, the relative moduli space must have an isometry group
larger than $SU(2)\times U(1)^{(k-1)}$, since at least some of the
extra zero modes are induces by the unbroken non-Abelian symmetry.
One might hope that this larger isometry group would act to simplify
the moduli space, just as the Taub-NUT manifold is much simpler than
the Atiyah-Hitchin.  Some partial progress in this direction has been
reported \cite{hollo}, and we are currently pursuing this direction.

Finally, one of the more compelling motivations for studying the low
energy dynamics of monopoles and dyons arises from the Montonen-Olive
duality conjecture \cite{dual}.  In certain supersymmetric Yang-Mills theories,
duality maps strongly coupled electric theories to weakly coupled
magnetic ones, thus enabling one to probe the nonperturbative nature
of strongly coupled Yang-Mills theories.  A notable example of this is
the $N=2$ supersymmetric gauge theories softly broken to $N=1$, where
confinement is explicitly realized through magnetic monopole
condensation \cite{seiberg}.

In order for such a duality to make sense, however, 
the spectrum of magnetically charged particles must be consistent with 
that predicted by the duality mapping of the electrically charged ones.
In $N=4$ supersymmetric Yang-Mills theories, duality relates 
each elementary massive vector meson of electric charge 
$\mbox{\boldmath $\gamma$}$ to 
a tower of dyons of magnetic charge $\mbox{\boldmath $\gamma$}^*$ among others,
where $\mbox{\boldmath $\gamma$}$ is any root of the gauge algebra.
The analysis of the classical solutions 
tells us, however, that a monopole based on a composite root is not a
fundamental entity, but rather corresponds to a mere coincidence 
point on a multi-monopole moduli space. Thus, it is imperative to see
if the quantization of the multi-monopole dynamics leads to a 
bound state that could conceivably be dual to the vector meson 
whenever $\mbox{\boldmath $\gamma$}$ is composite.

This program has recently been carried out in the simplest case, that
of a maximally broken $SU(3)$ gauge theory, by the authors  \cite{klee}, 
and also independently by Gauntlett and Lowe \cite{gaunt}. The most massive 
of the charged vector mesons in this theory is dual to a threshold bound state
with composite magnetic charge, which is realized as a unique normalizable
harmonic form on the Taub-NUT manifold.  Since, as we have seen above,
the Taub-NUT manifold is the universal relative moduli space for a
pair of interacting distinct fundamental monopoles, the same
two-monopole bound state immediately carries over to arbitrary gauge
groups \cite{klee}\cite{gaunt}.

It remains an open problem to find the bound states of more than two
fundamental monopoles. Now that there is a strong candidate for the
exact moduli space metric, however, it should be a matter of
differential calculus to search for harmonic forms and check if they
are suitably normalizable.  If this approach were to find the dyon
spectrum predicted by duality, it would provide further evidence in
support of proposed moduli space metric.

\vskip 1cm
\centerline{\large\bf Acknowledgments}
\vskip  5mm

We thank Jerome Gauntlett for useful conversations and Conor Houghton
for bringing Ref.~\cite{connell} to our attention.  This work was
supported in part by the U.S. Department of Energy. K.L. is supported
in part by the NSF Presidential Young Investigator program.

\end{document}